\documentstyle[twocolumn,aps,epsf]{revtex}
\newcommand{\be}{\begin{equation}}
\newcommand{\ee}{\end{equation}}
\newcommand{\bea}{\begin{eqnarray}}
\newcommand{\eea}{\end{eqnarray}}
\newcommand{\FN}[1]{$\typeout{Kommentar: #1}$}

\begin{document}
\bibliographystyle{prsty}
\title{Localization of  Waves without Bistability: 
Worms in Nematic Electroconvection}
\author{Hermann Riecke}
\address{Department of Engineering Sciences and Applied Mathematics\\
Northwestern University, Evanston, IL 60208, USA} 
\author{Glen D. Granzow}
\address{Department of Mathematics, Idaho State University, Pocatello, ID \FN{????} USA}
\maketitle

\begin{abstract}

A general localization mechanism for waves in dissipative systems is identified that
does not require the bistability of the basic state and the nonlinear plane-wave state. 
The mechanism explains the two-dimensional localized wave structures (`worms') 
that recently have been observed in experiments on electroconvection 
in nematic liquid crystals where the transition to extended waves is {\it supercritical}. 
The mechanism accounts for the propagation direction of 
the worms and certain aspects of their interaction. The dynamics of the localized waves
can be steady or irregular.


\end{abstract}

\centerline{submitted February 9, 1998}

 
A striking feature observed in a number of pattern-forming systems with large aspect
ratio is the spontaneous
localization or confinement of the pattern to a small part of the system although the
system is translationally invariant. Presumably the best studied structures of that 
type are the quasi-one-dimensional pulses of traveling waves that have been found 
in convection in binary mixtures (e.g. \cite{BeKo90}). They have been described theoretically
as perturbed solitons (e.g. \cite{ThFa88,Ri96}) and as bound 
pairs of fronts \cite{MaNe90,HeRi95}.  Other quasi-one-dimensional 
localized structures have been found in Taylor vortex flow \cite{WiAl92,GrSt97}, directional 
solidification \cite{SiBe88}, cellular flames \cite{BaMa94}, and in models for parametrically 
driven waves \cite{GrRi96}. 

In two dimensions localization appears to be harder to obtain; investigations of 
binary mixture convection have only led to long-lived, but eventually unstable wave pulses
\cite{LeBo93}. Only recently have truly stable two-dimensional localized 
structures been found in parametrically driven surface waves in granular media
and in a highly viscous fluid \cite{UmMe96}. 
In most systems the localized structures arise in a regime of bistability.
In such situations they can often be considered as a pair of fronts
 that are bound to each other by  dispersion \cite{MaNe90}, an
additional mode \cite{HeRi95}, or non-adiabatic effects 
\cite{BeSh88,SaBr96,Tu97,CrRi98}. 

Very recently \cite{DeAh96} 
a two-dimensional localized wave-state has been found in a system in which the
transition to the extended waves is {\it supercritical}, i.e. in which there exist no fronts
connecting the basic and the nonlinear state. This rules out the  
mechanisms of localization mentioned above. Thus, previous efforts to explain
these states turn out to be insufficient \cite{Tu97} and 
the origin of localization has remained
quite puzzling. The experiments have been performed in electroconvection
of nematic liquid crystals and due to their appearance the new localized states 
have been called `worms'. 

In this  paper we discuss a general mechanism that can lead to localization even if the bifurcation
to the extended waves is supercritical. It is based on the presence of an additional,
weakly damped field that is advected by the waves and that 
in turn affects their growth rate. The mechanism explains  
the shape of the worms, their direction of propagation and certain aspects of their interaction.

\begin{figure}[Htb]
  \begin{center}
    \leavevmode
    \epsfsize=0.3 \textwidth
    \epsffile{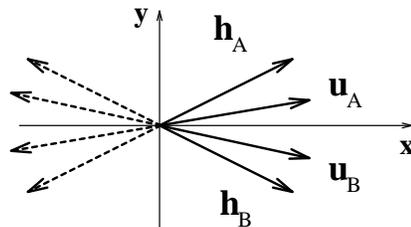}
    \caption{Sketch of the orientation of the group velocities
${\bf u_{A,B}}$ and  of the directions of advection ${\bf h_{A,B}}$ 
(cf. (\protect{\ref{e:cglA}}- \protect{\ref{e:cglC}}) below).
       }
    \label{f:zigzag}
  \end{center}
\end{figure}

Before the theoretical model is introduced a number of features of the experimentally
observed worms need to be discussed. 
Due to the preferred direction associated with the liquid
crystal the system has axial anisotropy. In the regime in question convection arises
in the form of waves that travel at a fixed angle relative to the axis of anisotropy. 
Due to reflection symmetry there are four such directions of propagation (left- and right-zig, and left- and right-zag) as indicated in fig.\ref{f:zigzag}.
The worms consist either of a combination of the left-traveling waves or of the
right-traveling waves.

 According to recent calcuations the transition to extended waves
is supercritical in this regime \cite{TrKr97}. 
This is consistent with the experiments, where for larger conductivity
of the liquid crystal
extended chaotic waves are observed to arise supercritically and no indication
of a change to subcritical behavior of these waves is seen \cite{DeAh96}.
Nevertheless, the worms arise already at parameter values at which
the state without convection is stable \cite{BiAh98}. 
In the $x$-direction the worms are quite
long and their length appears to be somewhat variable. In the $y$-direction,
however, they are very narrow and their width is fixed.
Unless perturbed, the worms travel only in the $x$-direction. 
In that direction
the envelope of the worm has a characteristic
shape: it rises rapidly to its maximum at one end and decays slowly over the
length of the worm. The worm travels toward the end with the large amplitude  
(its `head').

The worms occur very close to threshold and the bifurcation to the
extended traveling waves is supercritical \cite{TrKr97}. Therefore
 a set of coupled 
complex Ginzburg-Landau equations is considered. 
Since in a given developed worm 
a single set of zig- and zag-waves is observed \cite{DeAh96} 
we consider only these two
waves. 

Within the Ginzburg-Landau equations no sustained waves are possible below
threshold since all nonlinear terms are damping. We extend therefore these equations 
introduce an additional, weakly damped mode. The 
motivation for this mode arises from previous work in the context of
 binary-mixture convection where it was found by one of the authors
that pulses of traveling waves can arise
even for a supercritical bifurcation if the waves advect a mode that feeds back
into their growth rate \cite{Ri92a}. In addition, the weak-electrolyte model, which 
agrees quantitatively with the experiments with respect to the linear properties
\cite{DeTr96},
suggests that a charge-carrier mode becomes slow in the regime in which worms
appear \cite{TrKr97}. 

A minimal
model for the advection of a scalar mode by zig- and zag-waves is given by
\bea
\partial_t A&= &- {\bf u}_{A}\cdot \nabla A + \mu A + 
b_x \partial_x^2 A + b_y \partial_y^2 A + 2a\partial_{xy}^2A\label{e:cglA}\\
& &  + f C A -c|A|^2A - g|B|^2A, \nonumber\\
\partial_t B &=& - {\bf u}_{B} \cdot \nabla B + \mu B + 
b_x \partial_x^2 B + b_y \partial_y^2 B - 2a\partial_{xy}^2B\label{e:cglB}\\
& &  + f C B -c|B|^2B - g|A|^2B, \nonumber\\
\partial_t C &=& \delta \partial_x^2 C - \alpha C + 
{\bf h}_{A} \cdot \nabla |A|^2 
+ {\bf h}_{B} \cdot \nabla |B|^2. \label{e:cglC}
\eea
The equations for the complex wave amplitudes $A$ and $B$ are the usual complex
Ginzburg-Landau equations for oblique waves \cite{RiKr98}.
The equation for the scalar mode $C$ is obtained by
considering the currents ${\bf j}_{A}\equiv {\bf h}_{A} |A|^2$  and
${\bf j}_{B}\equiv {\bf h}_{B} |B|^2$ that are due to the advection by the respective
waves. In addition, damping and diffusion of the mode $C$ is allowed. 
In the following we first focus on the
effect of $C$ and will mostly consider the dispersionless case in which all 
coefficients are real. 

\begin{figure}[htb] 
\begin{picture}(420,300)(0,0)
\put(-10,180) {\includegraphics{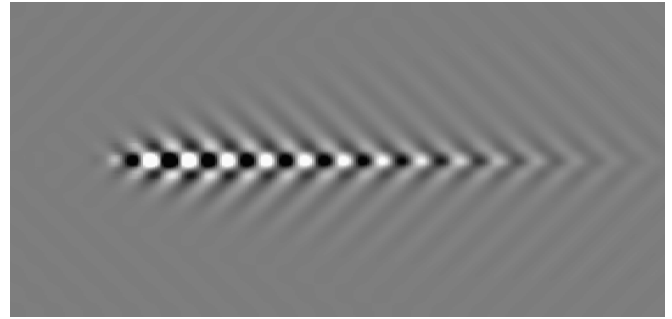}}
\put(-20,-70) {\includegraphics{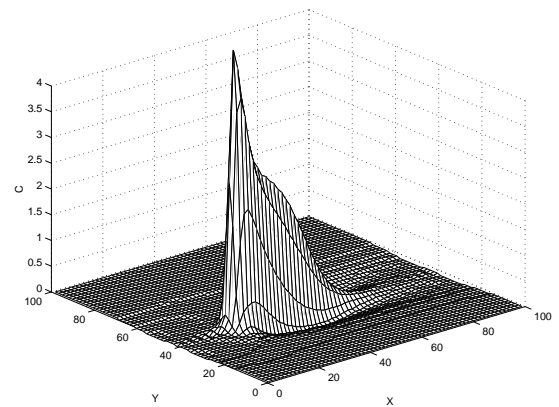}}
\end{picture}
\caption{Numerically determined worm-solution of 
({\protect{\ref{e:cglA}}-\protect{\ref{e:cglC}}}) for ${\bf u}_A=(0.05,0.75)$,
$\mu = -0.025$, $b_x=b_y=1$, $a=0$, $c=1$, $g=0.5$, $f=1$, $\alpha=-0.02$, 
$\delta=0.7$,
${\bf h}_A=(1,1)$. a) gray-scale plot of 
 $Ae^{i{\bf q}_A\cdot {\bf r}}+Be^{i{\bf q}_B\cdot {\bf r}}$. 
b) mode $C$. 
Note the depression of $C$ along the sides and the back of the worm. 
The worm travels to the left (toward its `head').
\protect{\label{f:worm}}
}
\end{figure}

Eqs.(\ref{e:cglA}-\ref{e:cglC}) are solved numerically using a pseudospectral 
code with an integrating-factor/Runge-Kutta time-stepping scheme.
Fig.\ref{f:worm} shows a worm solution obtained in this way.
For illustration purposes the top
part shows the real part of the quantity 
$Ae^{i{\bf q}_A\cdot {\bf r}}+Be^{i{\bf q}_B\cdot {\bf r}}$ which gives an indication 
of how
this solution would appear in experiments. The mode $C$ is shown in the bottom part.
As in the experiment the convective amplitude is large at one end of the
worm (`head') and decays slowly towards the other end. In agreement with the
experiments, the worm propagates toward their head. 

\begin{figure}[htb] 
\begin{picture}(420,120)(0,0)
\put(-20,-20) {\includegraphics{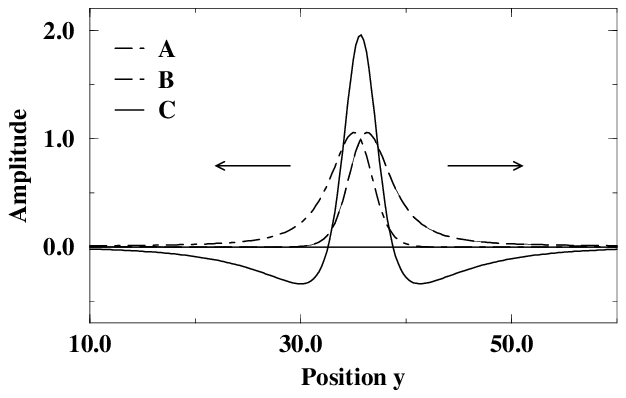}}
\end{picture}
\caption{Standing-wave pulse in one dimension ($y$-direction). 
Parameter values corresponding to fig.\protect{\ref{f:worm}}.
\protect{\label{f:swpuls}}
}
\end{figure}

To understand how this localized solution can arise already below the threshold $\mu =0$ although
the transition to extended waves is supercritical, two one-dimensional reductions 
of (\ref{e:cglA}-\ref{e:cglC}) are 
considered. The long and narrow shape of the worms suggests that the localization
mechanisms along and transverse to the worm differ from each other.
If one ignores the $x$-dependence, (\ref{e:cglA}-\ref{e:cglC}) reduce to two
equations describing standing waves coupled to $C$ and the worm is replaced by a 
stationary pulse
of standing waves. Such a solution is shown in fig.\ref{f:swpuls}. 
It exists because the $C$-mode enhances the growth rate in the center of the pulse and keeps the two components $A$ and $B$ together. Since the currents generating $C$ vanish if $|A|=|B|$, the pulse can only exist if the two traveling-wave components 
are shifted with respect to each other. 
The shift is due to the group velocity in the $y$-direction.
Thus, the worm disappears in 
a saddle-node bifurcation when the group velocity is reduced below a certain value. 
The standing-wave pulse is stable since the production of $C$ 
increases with the shift between $A$ and $B$.
 
For the understanding of the stability of the worm solution in fig.\ref{f:worm}
it is crucial to note that the standing-wave pulse shown in fig.\ref{f:swpuls} arises in a 
{\it subcritical bifurcation} although the bifurcation to the extended standing (or traveling)
waves is {\it supercritical}. Thus, the standing-wave pulse can exist already for $\mu<0$ and 
can coexist with the basic state $A=B=C=0$. 

The standing-wave pulse can  be considered
as a building block for the worm: in terms of a one-dimensional reduction along
 the $x$-direction the worm
can be seen as a pair of fronts connecting the basic state with the {\it coexisting standing-wave-pulse
state}. 
For stability these two fronts have to exhibit a repulsive interaction.  
With respect to the $x$-directon the zig and the zag making up the worms
are waves traveling in the same direction. Based on results for traveling-wave
pulses \cite{MaNe90,HeRi95,RiRa95} it is therefore expected that  
 the fronts can interact i) $via$ the wavenumber \cite{MaNe90}, 
which is driven by the dispersive terms, and ii)
$via$ the additional mode $C$ \cite{HeRi95,RiRa95}. 

If the interaction $via$ $C$ dominates, a simple connection
between the stability of the traveling-wave pulse (here, the worm) 
and its direction of propagation
emerges \cite{HeRi95}. As shown in fig.\ref{f:worm}b there is a 
large peak of $C$ at the head and a shallow depression of $C$ 
at the tail. Since positive (negative)  $C$ enhances (reduces) the local growth rate 
of the convective mode both fronts are pushed to the left by $C$. 
If the 
worm travels towards its head, the depression 
at the trailing front will be reduced  
by the remnants of the positive peak ahead of it. This is not the case for the leading peak.
Thus, the trailing front is pushed less to the left than the leading front amounting to a
repulsive interaction. For the traveling-wave pulses this interaction has been determined
previously in the limit of weak diffusion $\delta$ and damping $\alpha$ and for widely
separated fronts \cite{HeRi95}. 

For the present case of fronts in the standing-wave amplitude
the corresponding calculation would be considerably more involved. 
We expect, however, that the
same qualitative picture holds. This would imply that the worm is stable if
it travels towards its head and unstable otherwise \cite{HeRi95}.
Indeed the experimentally observed worms travel towards their head.
It should be noted that sufficiently strong dispersion can lead to additional stabilization
of the worms \cite{RiRa95}.
\FN{numerisch nachpruefen vergroessere u}

When increasing $\mu <0$ the width of the worm  and the maximal value of $C$
remain essentially unchanged, but the worm grows in length. 
Consequently, the spatial integral 
${\cal N} = \int |A|^2 dxdy$, which corresponds to a kind of Nusselt number,
increases smoothly except for a 
very small jump in ${\cal N}$ at the saddle-node bifurcation in which
the worm first appears. In a sufficiently large system, when the threshold $\mu=0$ is
surpassed a sequence of transitions to more than a single worm occurs. 
When the worms become long enough to span the whole system they 
loose their pronounced head structure and $C$ becomes independent of $x$. 

\begin{figure}[htb] 
\begin{picture}(420,150)(0,0)
\put(-55,-20) {\includegraphics{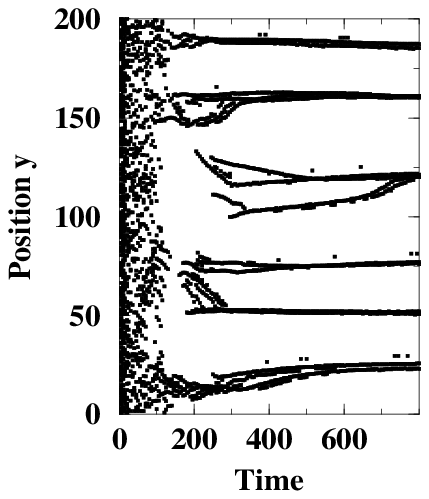}}
\put(115,34) {\includegraphics{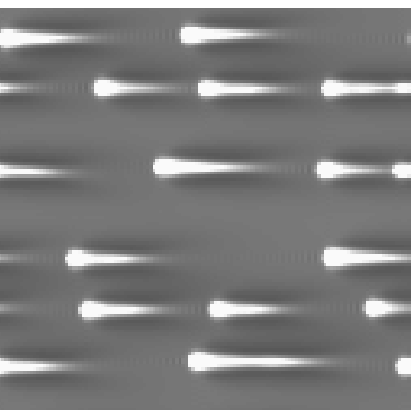}}
\end{picture}
\caption{Formation of worms from random initial conditions. Parameters as in fig.\protect{\ref{f:worm}} except for $\mu=0.05$ and ${\bf u}_A=(0.75,0.75)$.
a) $y$-location of the maxima in $C$ as a function of time. b) $C$-field for $t=800$. 
\protect{\label{f:ypos005}}
}
\end{figure}

In the experiments  the worms exhibit a typical spacing in the $y$-direction which decreases with
increasing applied voltage. To investigate this aspect within (\ref{e:cglA}-\ref{e:cglC}),
 the temporal evolution starting from random 
initial conditions is followed. 
Fig.\ref{f:ypos005}a gives the $y$-position
of the emerging worms  as a function of time for $\mu=0.05$. Fig.\ref{f:ypos005}b shows the
$C$-field of the worms at the final time of the run.  
As in the experiment, the worms approach roughly a
typical distance in the $y$-direction and almost travel on `tracks'. When $\mu$ is
increased the number of `tracks' is found to increase. 
\FN{ run fuer mu=0.2 machen und Anzahl nennen}
Fig.\ref{f:ypos005}b also demonstrates that the localization mechanism 
in the $y$-direction is much stronger than that in the $x$-direction: 
while all worms have essentially the same width they
vary substantially in their length. This is expected since the localization in the $x$-direction
is achieved $via$ the interaction of fronts, which is very weak and decays exponentially in space.
Thus, the timescales over which the worms reach their final length is much longer and perturbations
from other worms affect the length more strongly than the width.

So far the waves have been taken dispersionless. In the experiments it is found that
with increasing conductivity the worm regime crosses over to a regime characterized by 
bursts in the convective amplitude, i.e. patches or blobs of large amplitude appear and disappear
in an irregular fashion, and finally a regime of extended convection exhibiting spatio-temporal
chaos of patches of zigs, zags, and rectangles is reached \cite{DeAh96}. 
This suggests that dispersion 
becomes increasingly important along this path in parameter space and indeed the imaginary
parts of the coefficients in the Ginzburg-Landau equations for $A$ and $B$ (with $C$
eliminated) 
increase in this direction \cite{TrKr97}. 

\begin{figure}[htb] 
{\begin{picture}(420,150)(0,0)
\put(-40,-30) {\includegraphics{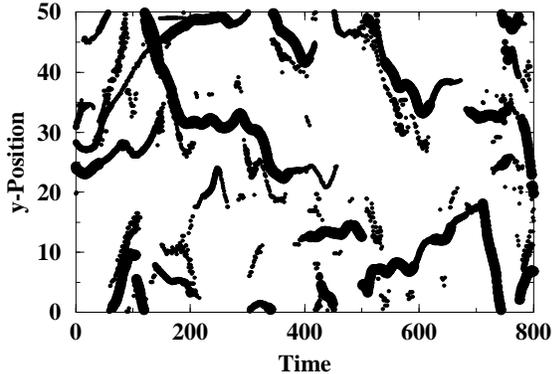}}
\end{picture}
\caption{Unsteady motion of worms. $y$-location as a function of time for 
${\bf u}_A=(0.05,0.05)$,
$\mu = 0.05$, $b_x=b_y=1-0.35i$, $a=0$, $c=1+0.5i$, $g=0.5+0.9i$, $f=1$, $\alpha=-0.05$, 
$\delta=0.4$, ${\bf h}_A=(1,1)$.
The size of the symbols indicates the height of the maximum of $C$. 
\protect{\label{f:yposdisp}}
}
}
\end{figure}

Fig.\ref{f:yposdisp} shows the result of a run with
intermediate values of the imaginary coefficients. 
The values have been chosen with some guidance from \cite{TrKr97}.
\FN{kann ich die fuer bursts nehmen von tREIBER}
In this regime 
the worms turn out to be unstable and start to travel not only in the $x$-direction 
but also in the $y$-direction. The motion in the $y$-direction 
is driven by imbalances between the amplitudes of the zig- and the
zag-component which lead to enhanced advection of $C$ towards one or the other
lateral sides of the worm. In addition,
the amplitude of the worms grows and decays as indicated by the size of the symbols
marking the $y$-position in fig.\ref{f:yposdisp}. 
These strong variations in the amplitude may be related
to the bursting seen in the experiments. 

In conclusion, we have introduced an extension to the complex Ginzburg-Landau equations for
zig- and zag-waves that leads in a simple way to localized, worm-like solutions although the 
bifurcation to extended waves is supercritical. 
The localization mechanisms differ in the $x$- and
the $y$-direction:  while the worm is spatially homoclinic in the $y$-direction, i.e.
there is only the basic state as a fixed point and the standing-wave pulse represents an 
excursion from it, it is heteroclinic in the $x$-direction, i.e. it connects the
basic state with a nonlinear periodic solution - the standing-wave pulse - $via$ two fronts. 
Within this framework the worms  travel towards their end with larger amplitude, 
in agreement with experimental observations. 
The observed typical spacing between the
worms in the $y$-direction is interpreted as the distance over which the advected field $C$ suppresses convection. If one assumes that the damping
of the additional mode increases with the conductivity the presented model captures
qualitatively the experimentally observed change from steady worms to 
extended spatio-temporal chaos $via$ an irregular bursting as the conductivity is 
increased. Extended spatio-temporal chaos is obtained when the damping
is large and the additional mode can be eliminated \cite{TrRi98}.

The localization presented here 
is similar in spirit to that found in parity-breaking bifurcations.
There the localized structures are drift waves embedded in a stationary, spatially
periodic state (rather than an unpatterned state) \cite{WiAl92,SiBe88,BaMa94} and  the additional field is the wavenumber 
of the underlying pattern \cite{RiPa92}. 

 HR grateful acknowledges discussions with L. Kramer, M. Treiber, Y. Tu, G. Ahlers, 
and M. Dennin. 
L. Kramer and M. Treiber kindly made their results available prior to publication. 
This research has
been supported by DOE under grant DE-FG02-92ER14303.
 

\end{document}